\documentstyle[aps,epsf,twocolumn]{revtex} 
\tightenlines 
\begin{document} 
\draft 
\title{
Magnon-Mediated Superconductivity in Itinerant Ferromagnets
         }
\author{Naoum Karchev\cite{byline}}
\address{
Department of Physics, University of Sofia, 1126 Sofia, Bulgaria
}
%
\maketitle

\begin{abstract}

The present paper discusses 
magnon-mediated superconductivity in ferromagnetic metals. The
mechanism explains in a natural way the fact that the
superconductivity in $UGe_2$, $ZrZn_2$ and
$URhGe$ is apparently confined to the ferromagnetic
phase.The order parameter is a spin anti-parallel component of
a spin-1 triplet with zero spin projection. The transverse spin
fluctuations are pair forming and the longitudinal ones are pair
breaking.The competition between magnons and paramagnons explains
the existence of two successive quantum phase transitions in
$UGe_2$, from ferromagnetism to ferromagnetic superconductivity,
and at higher pressure to paramagnetism. The maximum $T_{SC}$ results from
the suppression of the paramagnon contribution. To form a Cooper pair an
electron transfers from one Fermi surface to the other. As a result, the
onset
of superconductivity leads to the appearance of two Fermi
surfaces in each
of the spin up and spin down momentum
 distribution functions. This fact
explains the linear temperature
 dependence at low temperature of the specific
heat, and the experimental results for $UGe_2$.
 
\end{abstract}



The discovery of superconductivity in a single crystal of 
$UGe_2$\cite{mm1,mm2}, $ZrZn_2$\cite{mm3} and $URhGe$\cite{mm4}  
revived the interest in the coexistence of 
superconductivity and ferromagnetism.
The experiments indicate that in very pure systems, and at very low
temperature, ferromagnetism and superconductivity can coexist, with the same
electrons that cause the magnetism also responsible for the
superconductivity. The superconductivity is apparently 
confined to the
ferromagnetic phase. There are two successive
 quantum phase transitions in 
$UGe_2$\cite{mm1,mm2}, from ferromagnetism to
 ferromagnetic
superconductivity, and at higher pressure to paramagnetism. The specific
heat anomaly associated with the superconductivity appears to be absent and
the specific heat depends linearly on
 the temperature at low
temperature\cite{mm5,mm3,mm4}.
 
The paramagnon-mediated superconductivity
\cite{mm6,mm7} has long been
considered the most promising theory of
coexistence of superconductivity
and ferromagnetism.The order parameters are spin
parallel components of the spin triplet. The superconductivity in
$ZrZn_2$ was predicted. Nevertheless, the theory meets some
difficulties. It predicts that the superconductivity occurs in both
the ferromagnetic and the paramagnetic phases. A solution of the
problem was recently proposed. It has
been shown\cite{mm8} that the critical
temperature is much higher
in the ferromagnetic phase than in the paramagnetic
one due to the
coupling of the magnons to the longitudinal spin
fluctuations.
Alternatively, the superconductivity in the ferromagnetic
state of $ZrZn_2$ is explained in \cite{mm8a} as a result of an
exchange-type interaction between the magnetic moments of triplet-state
Cooper pairs and the ferromagnetic magnetization density in the
Ginzburg-Landau mean field theory. The Fay and Apple (FA) theory predicts that
spin up and spin down fermions
form Cooper pairs, and hence the specific heat
decreases
exponentially at low temperature. The phenomenological theories
\cite{mm9,mm10} circumvent the problem assuming that only
majority spin fermions form pairs, and hence the minority spin
fermions contribute the asymptotic of the specific heat. The coefficient
$\gamma=\frac CT$ is twice smaller in the
superconducting phase, which closely matches the experiments with
$URhGe$\cite{mm4}. But the assumption is quite questionable when the
magnetization approaches zero. The superconducting critical temperature in
(FA) theory increases when the magnetization decreases and very close to the
quantum critical point falls down rapidly.  It has recently been the subject
of controversial debate. It is obtained in\cite{mm11}, by means of a more
complete Eliashberg treatment, that the transition temperature is nonzero at
the critical point. In \cite{mm12},  however, the authors have shown that the
reduction of quasiparticle coherence and life-time due to spin fluctuations is
the pair-breaking  process which leads to a rapid reduction of the
superconducting critical temperature near the quantum critical point. 
Finally, recent studies of polycrystalline samples of UGe2 show that T-P phase diagram
is very similar to those of single-crystal specimens of UGe2 \cite{mm13}. These
findings suggest that the superconductivity in UGe2 is relatively 
insensitive to the presence of impurities and defects which excludes
the spin parallel pairing.

In the present paper an itinerant system is considered in which the
spin-$\frac 12$ fermions responsible for the ferromagnetism are the same
quasiparticles which form the Cooper pairs. The interaction of quasiparticles
$c_{\sigma}(\vec x)(c^+_{\sigma}(\vec x))$ with spin fluctuations has the form
\FL
\begin{equation} 
H_{\text{s-f}}=J\int d^3x \,c^+(\vec x)\frac {\vec \tau}{2}c(\vec x)\cdot 
\vec {M}(\vec x)
\label{mm1a} 
\end{equation}
where the transverse spin fluctuations are described by
magnons  $M_1(\vec x)+iM_2(\vec x)=\sqrt{2M}a(\vec x)$,
$M_1(\vec x)-iM_2(\vec x)=\sqrt{2M}a^+(\vec x)$ and the longitudinal spin
fluctuations by paramagnons $M_3(\vec x)-M=\varphi(\vec x)$. $M$ is zero
temperature dimensionless magnetization of the system per lattice site. The
magnon's dispersion is $\omega(\vec k)=\rho\vec k^2$ where the spin stiffness
constant is proportional to $M$ ($\rho=M\rho_0$), and the paramagnon
propagator is
\FL
\begin{equation} 
D_{\text{pm}}=\frac {1}{r-i\frac {\omega}{|\vec p|}+b\vec p^2}.
\label{mm1b}
\end{equation} 
The
parameter $r$ is the inverse static longitudinal magnetic susceptibility,
which measures the deviation from quantum critical point. The constants
$J,\rho_0$ and $b$ are phenomenological ones subject to some relations.

Integrating out the spin fluctuations one obtains an effective four fermion
theory which can be written as a sum of four terms. Three of them describe the
interaction of the components of spin-1 composite fields $(\uparrow\uparrow,
\uparrow\downarrow+\downarrow\uparrow, \downarrow\downarrow)$ which have a
projection of spin 1,0 and -1 respectively. The fourth term describes the
interaction of the spin singlet composite fields
$\uparrow\downarrow-\downarrow\uparrow$. The spin singlet fields' interaction
is repulsive and does not contribute to the superconductivity \cite{mm13a}. The
spin parallel fields' interactions are due to the exchange of paramagnons and
do not contribute to the magnon-mediated superconductivity. The relevant
interaction is that of the $\uparrow\downarrow+\downarrow\uparrow$ fields. In
static approximation, the Hamiltonian of interaction is 
\begin{eqnarray}
\FL
& & H_{\text int} = -\frac {J^2}{16}\int\prod\limits_{i}\frac
{d^3k_i}{(2\pi)^3}
\left[c_{\uparrow}^+(\vec k_1)c_{\downarrow}^+(\vec k_2)+
c_{\downarrow}^+(\vec k_1)c_{\uparrow}^+(\vec k_2)\right] \nonumber \\
\\
\label{mm1}  
& & \left[c_{\uparrow}(\vec k_2-\vec k_3)c_{\downarrow}(\vec k_1+\vec k_3)+
c_{\downarrow}(\vec k_2-\vec k_3)c_{\uparrow}(\vec k_1+\vec k_3)\right]
V(\vec k_3) \nonumber 
\end{eqnarray}
where the potential
\FL
\begin{equation}
V(\vec k)=\frac {2M}{\rho\vec k^2}-\frac {1}{r+b\vec k^2}
\label{mm2}
\end{equation}
has an attracting part due to exchange of magnons and a repulsive part due to
exchange of paramagnons. The effective Hamiltonian of the system is
\FL
\begin{equation}
H_{\text eff}=H_0+H_{\text int}
\label{mm21}
\end{equation}
where  $H_0$ is the Hamiltonian of the free spin
up and spin down fermions with dispersions
\FL
\begin{eqnarray} 
\epsilon_{\uparrow}(\vec k) & =
& \frac {\vec k^2}{2m}-\mu-\frac {JM}{2},
\nonumber \\  
\epsilon_{\downarrow}(\vec k) & = & \frac {\vec k^2}{2m}-\mu +\frac {JM}{2}
\label{mm22}
\end{eqnarray} 
One can obtain the gap equation following the standard technique.
To ensure that the
fermions which form Cooper pairs are the same as those responsible for
spontaneous magnetization, one has to consider the equation for the
magnetization
\FL
\begin{equation} 
M=\frac 12 <c^+_{\uparrow}c_{\uparrow}-c^+_{\downarrow}c_{\downarrow}>
\label{mm2a}
\end{equation}
as well.
Then the system of equations for the gap and for the magnetization determines
the phase where the superconductivity and the ferromagnetism coexist.The system 
can be written in terms of Bogoliubov excitations, which have
the following dispersions relations:
\FL
\begin{eqnarray}
& & E_1(\vec k) = -\frac
{JM}{2}-\sqrt{\epsilon^2(\vec k)+|\Delta(\vec
k)|^2} \nonumber \\ 
& & E_2(\vec k) = \frac
{JM}{2}-\sqrt{\epsilon^2(\vec
k)+|\Delta(\vec k)|^2}  
\label{mm2b}
\end{eqnarray}
where $\Delta(\vec k)$
is the gap, and $\epsilon(\vec k)=\frac {\vec k^2}{2m}-\mu$.

At zero temperature the equations take the form
\FL
\begin{eqnarray}
M & = & \frac 12\int\frac {d^3k}{(2\pi)^3}\left[1-\Theta (-E_2(\vec k))\right]
\label{mm3}\\
\Delta(\vec p) & = & \frac {J^2}{8}\int\frac {d^3k}{(2\pi)^3}\,\frac {V(\vec
p-\vec k)\,\,\Theta (-E_2(\vec k))}{\sqrt {\epsilon^2(\vec k)+|\Delta(\vec
k)|^2}}\,\Delta (\vec k)
\label{mm4}
\end{eqnarray}

The gap is an antisymmetric function $\Delta (-\vec k)=-\Delta (\vec k)$, so
that the expansion in terms of spherical harmonics $Y_{lm}(\Omega_{\vec k})$
contains only terms with odd $l$. I assume that the component with $l=1$ and
$m=0$ is nonzero and the other ones are zero 
\FL
\begin{equation}
\Delta (\vec k)=\Delta_{10}(k)\sqrt {\frac {3}{4\pi}}\cos\theta.
\label{mm4a}
\end{equation}
  Expending
the potential in terms of Legendre polynomial $P_l$ one obtains that only the
component with $l=1$ contributes the gap equation. The potential $V_1(p,k)$
has the form, \FL
\begin{eqnarray}
V_{1}(p,k) & = & \frac {3M}{\rho}\left[\frac
{p^2+k^2}{4p^2k^2}\ln\left(\frac {p+k}{p-k}\right)^2-\frac{1}{pk}\right]
- \nonumber \\
& & 
\frac {3M}{\rho}\beta \left[\frac {p^2+k^2}{4p^2k^2}\ln\frac
{r'+(p+k)^2}{r'+(p-k)^2}\,-\,\frac {1}{pk}\right], 
\label{mm5}
\end{eqnarray}
where $\frac {3M}{\rho}=\frac {3}{\rho_0}$, $\beta=\frac {\rho}{2Mb}=\frac
{\rho_0}{2b}>1$ and $r'=\frac {r}{b}<<1$. A straightforward analysis 
shows that for a fixed $p$ , the potential is positive when $k$
runs an interval around $p$ $(p-\Lambda,p+\Lambda)$,
where $\Lambda$ is approximately independent on $p$. 
In order to allow for an explicit analytic solution, I introduce further
simplifying assumptions by neglecting the dependence of $\Delta_{10}(k)$
on $k$ ($\Delta_{10}(k)=\Delta_{10}(p_f)=\Delta$) and setting
$V_1(p_f,k)$ equal to a constant $V_1$ within interval
$(p_f-\Lambda,p_f+\Lambda)$ and zero elsewhere. The system of equations
(\ref{mm3},\ref{mm4}) 
is then reduced to the system
\FL
\begin{eqnarray}
M & = & \frac
{1}{8\pi^2}\int\limits_0^{\infty}dkk^2\int\limits_{-1}^{1}dt[1-\Theta(-E_2(k,t))]
\label{mm6}\\
\Delta & = & \frac {J^2V_1}{32\pi^2}\int\limits_{p_f-\Lambda}^{p_f+\Lambda} dk
k^2\int\limits_{-1}^{1}dt\,t^2\frac
{\Theta(-E_2(k,t))}{\sqrt{\epsilon^2(k)+\frac {3}{4\pi}t^2\Delta^2}}
\Delta
\label{mm7}
\end{eqnarray}
where $t=\cos\theta$.

One looks for a solution of the system which satisfies $\sqrt {\frac
{3}{\pi}}\Delta<JM$. Then the equation $E_2(k,t)=0$ defines the
Fermi surfaces
\FL 
\begin{equation}
k^{\pm}_{f}=\sqrt {p^2_{f}\pm m\sqrt{J^2M^2-\frac
{3}{\pi}t^2\Delta^2}}\,\,,\,\,\, p_{f}=\sqrt {2\mu m}
\label{mm8}
\end{equation} 
The domain between the Fermi surfaces contributes to the magnetization $M$ in
Eq.(\ref{mm6}), but it is cut out from the domain of integration in the gap
equation Eq.(\ref{mm7}). One is primarily interested in determining at what
magnetization a superconductivity exists. When the magnetization increases, the
domain of integration in the gap equation decreases. Near the quantum
critical point the size of the gap is small, and hence the linearized gap
equation can be considered. Then it is easy to obtain the critical value of
the magnetization $M_{SC}$
\FL
\begin{equation}
M_{SC}=\frac {4p^{2}_{f}}{mJ}\left(1+\frac
{4p^2_{f}-\Lambda^2}{\Lambda^2}\exp\left[\frac {64\pi^2}{3J^2 V_1 m p_{f}}
\right]\right)^{-\frac 12} 
\label{mm9}
\end{equation}

Near the second quantum phase transition, when the magnetization approaches
zero, the domain between the Fermi surfaces decreases. One can approximate the
equation for magnetization Eq.(\ref{mm6}) substituting $k^{\pm}_{f}$ from
Eq.(\ref{mm8}) in the
the difference $(k^{+}_{f})^2-(k^{-}_{f})^2$ and setting
$k^{\pm}_{f}=p_{f}$ elsewhere. Then, in this approximation, the
magnetization is linear in $\Delta$, namely
\FL
\begin{equation}
\Delta =\sqrt {\frac {\pi}{3}}J\kappa M
\label{mm10}
\end{equation}
where $\kappa$ runs the interval $(0,1)$, and satisfies the equation
\FL
\begin{equation}
\kappa\sqrt {1-\kappa^2}+\arcsin\kappa\,=\frac {8\pi^2}{mp_{f}J}
\label{mm11}
\end{equation}
The Eq.(\ref{mm11}) has a solution if $mp_{f}J>16\pi$. Substituting $M$ from
Eq.(\ref{mm10}) in Eq.(\ref{mm7}), one arrives at an equation for the gap. This
equation can be solved in a standard way and the solution is
\FL
\begin{eqnarray}
\Delta & = & \sqrt {\frac {16\pi}{3}}\frac {\Lambda p_{f}\kappa}{m}
\exp\left[-\frac32
I(\kappa)-\frac {24\pi^2}{J^2 V_1 m p_{f}}\right]
\label{mm12} \\
I(\kappa) & = & \int\limits_{-1}^{1}dt t^2 \ln\left(1+\sqrt
{1-\kappa^2 t^2}\right) \nonumber
\end{eqnarray}
Eqs (\ref{mm10},\ref{mm11},\ref{mm12}) are the solution of the system
Eqs.(\ref{mm6},\ref{mm7}) near the quantum transition to paramagnetism. 

When superconductivity and ferromagnetism coexist, the momentum
distribution functions $n^{\uparrow}(p,t)$ and $n^{\downarrow}(p,t)$ of the
spin-up and spin-down quasiparticles have two Fermi surfaces each. One can
write them in terms of the distribution functions of the Bogoliubov fermions
\FL
\begin{eqnarray}
n^{\uparrow}(p,t) & = & u^2(p,t)n_1(p,t)+v^2(p,t)n_2(p,t)
\label{mm13} \\
n^{\downarrow}(p,t) & = & u^2(p,t)(1-n_1(p,t))+v^2(p,t)(1-n_2(p,t))
\nonumber
\end{eqnarray}
where $u(p,t)$ and $v(p,t)$ are the coefficients in the Bogoliubov
transformation. At zero temperature $n_1(p,t)=1$,
$n_2(p,t)=\Theta(-E_2(p,t))$, and the Fermi surfaces Eq.(\ref{mm8}) manifest
themselves both in the spin-up and spin-down momentum distribution functions.
The functions are depicted in Fig.1 and Fig.2.

\begin{figure}[h] 
\vspace{0.3cm} 
\epsfxsize=7.0cm 
\hspace*{0.2cm} 
\epsfbox{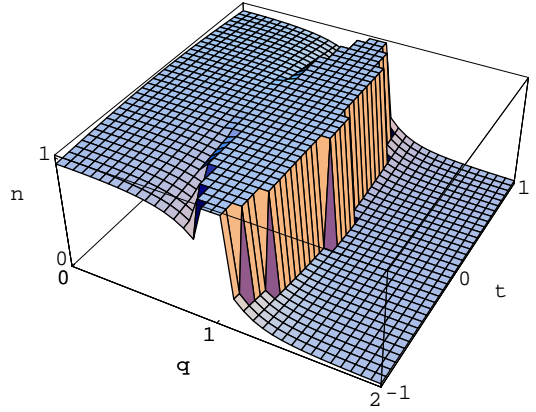} 
\caption{The zero temperature momentum distribution $n$, for spin up 
fermions, as a function of $q=\frac {p}{p_{f}}$ and $t=\cos\theta$.} 
\label{fig1} 
\end{figure} 

\begin{figure}[h] 
\vspace{0.3cm} 
\epsfxsize=7.0cm 
\hspace*{0.2cm} 
\epsfbox{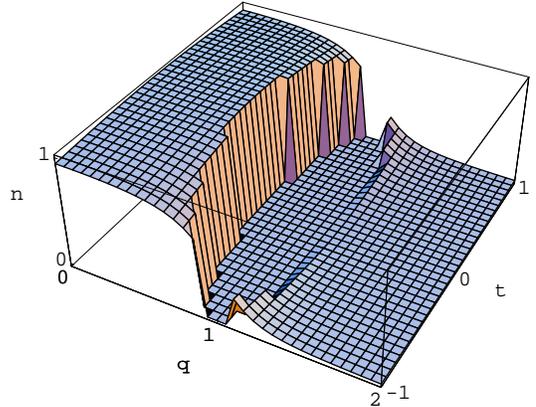} 
\caption{The zero temperature momentum distribution $n$, for spin down
fermions, as a function of $q=\frac {p}{p_{f}}$ and $t=\cos\theta$.} 
\label{fig2} 
\end{figure}

The two Fermi surfaces are necessary for the existence of itinerant
ferromagnetism , and explain the mechanism of Cooper pairing. In
the ferromagnetic phase $n^{\uparrow}$ and $n^{\downarrow}$ have different
(majority and minority) Fermi surfaces. To form a spin anti-parallel Cooper
pair, the fermion has to transfer from one Fermi surface to the other. If the
value of the momentum of the emitted or absorbed magnon lies in the domain
where
 the effective potential is attracting but outside the domain between
the
 two Fermi surfaces, fermions with opposite spins form a Cooper pair. As a
result, the onset of superconductivity is accompanied by the appearance of a
second Fermi surface in each of the spin-up and spin-down momentum
distribution functions. 

The existence of the two Fermi surfaces also explains the linear dependence
of the specific heat at low temperatures as opposed to the exponential
decrease of the specific heat in the BCS theory:
 
\FL
\begin{equation}
\frac {C}{T}\,=\,\frac {2\pi^2}{3}\left(N^+(0)+N^-(0)\right)
\label{mm14}
\end{equation}
Here $N^{\pm}(0)$ are the density of states on the Fermi surfaces.
One ca rewrite the $\gamma=\frac {C}{T}$ constant in terms of Elliptic
Integral of the second kind $E(\alpha,x)$
\FL
\begin{eqnarray}
\gamma\,=\,\frac {m p_{f}}{3\kappa} & &
\left[(1+s)^{\frac 12} E(\frac 12 \arcsin\kappa,\frac
{2s}{s+1})+\right. \nonumber  \\  
& & \left. (1-s)^{\frac 12} E(\frac 12 \arcsin\kappa,\frac
{2s}{s-1})\right].
\label{mm15}
\end{eqnarray}
where $s=\frac {JMm}{p^2_{f}}<1$ and 
$\kappa=\sqrt {\frac {3}{\pi}}\frac {\Delta}{JM}$.
The Eq.(\ref{mm15}) shows that in the ferromagnetic phase
($\Delta=0$) the specific heat constant $\gamma$  is smaller than in
the superconducting one, which closely matches the
experiments with $UGe_2$\cite{mm5}.
Important point is that UGe2 is an
anisotropic ferromagnet and hence the magnon has a gap which changes the
potential Eq.(\ref{mm5}). The physical consequence of the change is that the
superconductivity disappears before the quantum phase transition from
ferromagnetism to paramagnetism (see \cite{mm5,mm13}). The distance between
these two points
 depends on the anisotropy. 

The presence of an additional phase line $T_{x}$ and the correlation
between a transition at $T_{x}$ and the appearance of superconductivity in
$UGe_2$ has been proved \cite{mm2,mm5}. It lies entirely within the
ferromagnetic phase and is suggested by a strong anomaly in the resistivity at
a temperature $T_{x}$. The maximum transition temperature for
superconductivity is near the pressure $P_{x}$ where $T_{x}$ vanishes.
The authors have assumed that superconductivity is mediated by
fluctuations associated with a (hypothetical) second order critical point
$P_{x}$, with an unidentified order parameter.

Experimental measurement of ac magnetic susceptibility as a function of the
temperature indicates a peak at ferromagnetic Curie temperature, as usual.
But the peak is substantially damped at a pressure near the maximal
superconducting critical temperature \cite{mm14}. The
suppression of the peak can be understood as a suppression of the
paramagnon contribution, which in turn means suppression of pair-breaking
effects and hence higher superconducting critical temperature. So the proposed
model  of magnon mediated superconductivity complemented by the experimental
results explains, at least qualitatively, the superconductivity in $UGe_2$
without invoking an additional phase transition.

The proposed model of ferromagnetic superconductivity differs from the
models discussed in \cite{mm6,mm7,mm8,mm8a,mm9,mm10} in two ways. First, the
superconductivity is due to the exchange of magnons, and paramagnons have
pair-breaking effect. Second, the order parameter is a
spin antiparallel component of a spin triplet with zero spin projection.
The existence of two Fermi surfaces in each of the spin-up and spin-down
momentum distribution functions is a generic property of a ferromagnetic
superconductivity with spin anti-parallel pairing (see also\cite{mm15}). They
lead to a linear temperature dependence of the specific heat at low
temperature. But the experimental result has an alternative theoretical
explanation in \cite{mm4,mm9,mm10}. So, one needs an experiment which proves
undoubtedly the existence of the predicted Fermi surfaces.

The author would like to thank C. Pfleiderer and R.Rashkov for valuable discussions.

\end{document}